\documentclass{easychair}

\usepackage{doc}
\usepackage{etex}
\usepackage[utf8]{inputenc}
\usepackage{lmodern}
\usepackage{hyphsubst}
\usepackage[english]{babel}
\usepackage{csquotes}
\usepackage{textcomp}
\usepackage{enumerate}
\usepackage{microtype}
\usepackage{listings}
\usepackage{graphicx}
\usepackage{subcaption}
\usepackage{bbold}
\usepackage{float}
\usepackage[binary-units=true]{siunitx}
	\DeclareSIUnit{\belmilliwatt}{Bm}
	\DeclareSIUnit{\dBm}{\deci\belmilliwatt}
	\DeclareSIUnit{\beliso}{Bi}	
	\DeclareSIUnit{\dBi}{\deci\beliso}	
\usepackage{graphicx}
\usepackage{amsmath,amssymb,amsfonts}
\usepackage[nolist]{acronym}
\usepackage{tikz}
\usetikzlibrary{automata,positioning,fit}
\usepackage{algorithm}
\usepackage[noend]{algpseudocode}
\usepackage{todonotes}
\usepackage{bbold}
\usepackage{color}
\usepackage[backend=biber, sorting=none, block=ragged]{biblatex}
\usepackage{adjustbox}
\renewcommand{\quote}[1]{``#1''}

\begin{acronym}

\acro{4g}[4G]{Fourth Generation}
\acro{5g}[5G]{Fifth Generation}
\acro{6lowpan}[6LoWPAN]{IPv6 over Low-Power Wireless Personal Area Networks}
\acro{6lowpan-nd}[6LoWPAN-ND]{IPv6 over Low-Power Wireless Personal Area Networks with Neighbor Discovery}
\acro{6p}[6P]{6top Protocol}
\acro{6tisch}[6TiSCH]{IPv6 over the TSCH mode of IEEE 802.15.4e}
\acro{6top}[6top]{6TiSCH Operation Sublayer}
\acro{a2a}[A2A]{Air-to-Air}
\acro{a2g}[A2G]{Air-to-Ground}
\acro{aanet}[AANET]{Aeronautical Ad-hoc Network}
\acro{ac}[AC]{aircraft}
\acro{ack}[ACK]{acknowledgement}
\acro{adsb}[ADS-B]{Automatic Dependent Surveillance - Broadcast}
\acro{ae}[AE]{Auto Encoder}
\acro{afdx}[AFDX]{Avionic Full Duplex}
\acro{afh}[AFH]{Adaptive Frequency Hopping Spread Spectrum}
\acro{ai}[AI]{Artificial Intelligence}
\acro{ais}[AIS]{Automatic Identification System}
\acro{ann}[ANN]{Artificial Neural Network}
\acro{aodv}[AODV]{Ad hoc On-Demand Distance Vector}
\acro{aodvv2}[AODVv2]{Ad Hoc On-demand Distance Vector Version 2}
\acro{ap}[AP]{Access Point}
\acro{arns}[ARNS]{Aeronautical Radionavigation Service}
\acro{arq}[ARQ]{Automatic Repeat Request}
\acro{asn}[ASN]{Absolute Slot Number}
\acro{atc}[ATC]{Air Traffic Control}
\acro{atm}[ATM]{Air Traffic Management}
\acro{awgn}[AWGN]{Additive White Gaussian Noise}
\acro{bc}[BC]{Broadcast Channel}
\acro{ber}[BER]{Bit Error Rate}
\acro{bmwi}[BMWi]{Bundesministerium für Wirtschaft und Energie}
\acro{ban}[BAN]{Body Area Network}
\acro{cca}[CCA]{Clear Channel Assessment}
\acro{ccdf}[CCDF]{Complementary Cumulative Distribution Function}
\acro{cdf}[CDF]{Cumulative Distribution Function}
\acro{cnn}[CNN]{Convolutional Neural Network}
\acro{cr}[CR]{Cognitive Radio}
\acro{crn}[CRN]{Cognitive Radio Network}
\acro{csmaca}[CSMA/CA]{Carrier Sense Multiple Access/Collision Avoidance}
\acro{csma}[CSMA]{Carrier Sense Multiple Access}
\acro{cstdma}[CSTDMA]{Carrier Sense TDMA}
\acro{css}[CSS]{Cooperative Spectrum Sensing}
\acro{d2d}[D2D]{Device-to-Device}
\acro{darpa}[DARPA]{U.S. Defense Advanced Research Projects Agency}
\acro{dbn}[DBN]{Deep Belief Network}
\acro{dfg}[DFG]{Deutsche Forschungsgemeinschaft}
\acro{dl}[DL]{downlink}
\acro{dllayer}[DL]{Data Link}
\acro{dme}[DME]{Distance Measuring Equipment}
\acro{dnn}[DNN]{Deep Neural Network}
\acro{dodag}[DODAG]{Destination Oriented Directed Acyclic Graph}
\acro{drl}[DRL]{Deep Reinforcement Learning}
\acro{dsa}[DSA]{Dynamic Spectrum Access}
\acro{dsme}[DSME]{Deterministic and Synchronous Multi-channel Extension}
\acro{dsss}[DSSS]{Direct-Sequence Spread Spectrum}
\acro{dtmc}[DTMC]{discrete-time Markov chain}
\acro{dtx}[DTX]{discontinuous reception}
\acro{dqn}[DQN]{Deep $Q$-Network}
\acro{eb}[EB]{Enhanced Beacon}
\acro{ed}[ED]{Energy Detection}
\acro{embb}[eMBB]{Enhanced Mobile Broadband}
\acro{enb}[eNB]{evolved NodeB}
\acro{etx}[ETX]{Expected Transmission Count}
\acro{fanet}[FANET]{Flying ad hoc Network}
\acro{fdm}[FDM]{Frequency Division Multiplexing}
\acro{fdma}[FDMA]{Frequency-Division Multiple Access}
\acro{fdr}[FDR]{Frame Delivery Ratio}
\acro{fec}[FEC]{Forward Error Correction}
\acro{fft}[FFT]{Fast Fourier Transformation}
\acro{fhss}[FHSS]{Frequency Hopping Spread Spectrum}
\acro{fl}[FL]{Flight Level}
\acro{flink}[FL]{Forward Link}
\acro{fmcw}[FMCW]{Frequency Modulated Continuous Wave}
\acro{frr}[FRR]{frequency resource reuse}
\acro{fspl}[FSPL]{Free-Space Path Loss}
\acro{gan}[GAN]{Generative Adversarial Network}
\acro{gd}[GD]{Gradient Descent}
\acro{ge}[GE]{Gilbert-Elliot}
\acro{gps}[GPS]{Global Positioning System}
\acro{gpsr}[GPSR]{Greedy Perimeter Stateless Routing}
\acro{harq}[HARQ]{Hybrid Automatic Repeat Request}
\acro{hmm}[HMM]{Hidden Markov Model}
\acro{hs}[HS]{Hopping Sequence}
\acro{icao}[ICAO]{International Civil Aviation Organization}
\acro{idm}[IDM]{Interference Detection Mechanism}
\acro{ieee}[IEEE]{Institute of Electrical and Electronics Engineers}
\acro{ie}[IE]{Information Element}
\acro{iesg}[IESG]{Internet Engineering Steering Group}
\acro{ietf}[IETF]{Internet Engineering Task Force}
\acro{intairnet}[IntAirNet]{Inter Aircraft Network}
\acro{ioe}[IoE]{Internet of Everything}
\acro{iot}[IoT]{Internet of Things}
\acro{iiot}[IIoT]{Industrial Internet of Things}
\acro{iov}[IoV]{Internet of Vehicles}
\acro{ip}[IP]{Internet Protocol}
\acro{ipm}[IPM]{Interference Prediction Mechanism}
\acro{irtf}[IRTF]{Internet Research Task Force}
\acro{ism}[ISM]{industrial, scientific and medical}
\acro{itu}[ITU]{International Telecommunication Union}
\acro{itur}[ITU-R]{International Telecommunication Union Radiocommunication Sector}
\acro{jtids}[JTIDS]{Joint Tactical Information Distribution System}
\acro{kpi}[KPI]{Key Performance Indicator}
\acro{laa}[LAA]{License-Assisted Access}
\acro{ldacs}[LDACS]{$L$-band Digital Aeronautical Communications System}
\acro{ldg}[LDG]{Landing Gear}
\acro{lln}[LLN]{Low-power and Lossy Network}
\acro{lorawan}[LoRaWAN]{Long Range Wide Area Network}
\acro{lpwan}[LPWAN]{Low-Power Wide Area Network}
\acro{lqi}[LQI]{Link Quality Indicator}
\acro{lre}[LRE]{Limited Relative Error}
\acro{lstm}[LSTM]{Long Short Term Memory}
\acro{lte}[LTE]{Long Term Evolution}
\acro{lte-u}[LTE-U]{Long Term Evolution Unlicensed}
\acro{lufo}[LuFo]{Luftfahrtforschung}
\acro{mac}[MAC]{Medium Access Control}
\acro{amanet}[avionic MANET]{Avionic Mobile Ad-hoc Network}
\acro{manet}[MANET]{Mobile Ad-hoc Network}
\acro{mc}[MC]{Markov Chain}
\acro{mcs}[MCS]{Modulation and Coding Scheme}
\acro{mdp}[MDP]{Markov Decision Process}
\acro{ml}[ML]{Machine Learning}
\acro{mlp}[MLP]{Multi-Layer Perceptron}
\acro{mmtc}[mMTC]{massive Machine-Type Communications}
\acro{mos}[MOS]{Mean Opinion Score}
\acro{mse}[MSE]{Mean Squared Error}
\acro{msfg}[MSFG]{Matrix Signal Flow Graph}
\acro{msf}[MSF]{Minimal Scheduling Function}
\acro{msk}[MSK]{Minimum-Shift Keying}
\acro{mso}[MSO]{Message Start Opportunity}
\acro{mtsch}[MTSCH]{Mobility-aware Time Slotted Channel Hopping}
\acro{mtu}[MTU]{Maximum Transmission Unit}
\acro{nack}[NACK]{negative acknowledgment}
\acro{nac}[NAC]{North Atlantic Corridor}
\acro{nato}[NATO]{North Atlantic Treaty Organization}
\acro{nbiot}[NB-IoT]{Narrowband Internet of Things}
\acro{nextgen}[NextGen]{Next Generation Air Transportation System}
\acro{nic}[NIC]{Network Interface Card}
\acro{nice}[NICE]{Non-Intrusive Channel quality Estimation}
\acro{oca}[OCA]{Oceanic Control Area}
\acro{ofdm}[OFDM]{Orthogonal Frequency Division Multiplexing}
\acro{olsr}[OLSR]{Open Link State Routing}
\acro{oqpsk}[OQPSK]{Offset Quadrature Phase-Shift Keying}
\acro{osi}[OSI]{Open Systems Interconnection}
\acro{p2p}[P2P]{Point-to-Point}
\acro{pan}[PAN]{Personal Area Network}
\acro{pca}[PCA]{Priority Channel Access}
\acro{pdf}[PDF]{Probability Density Function}
\acro{pdr}[PDR]{Packet Delivery Ratio}
\acro{pdu}[PDU]{Protocol Data Unit}
\acro{per}[PER]{Packet Error Rate}
\acro{pf}[PF]{Proportional Fair}
\acro{pgf}[PGF]{Probability Generating Function}
\acro{phy}[PHY]{Physical}
\acro{pib}[PIB]{PAN Information Base}
\acro{pl}[PL]{Packet Loss}
\acro{pmf}[PMF]{Probability Mass Function}
\acro{ppp}[PPP]{Poisson Point Process}
\acro{prb}[PRB]{Physical Resource Block}
\acro{prr}[PRR]{Packet Reception Rate}
\acro{psd}[PSD]{Power Spectral Density}
\acro{pu}[PU]{Primary User}
\acro{qam}[QAM]{Quadrature Amplitude Modulation}
\acro{qos}[QoS]{Quality of Service}
\acro{ra}[RA]{Radio Altimeter}
\acro{rbm}[RBM]{Restricted Boltzmann Machine}
\acro{rb}[RB]{resource block}
\acro{rch}[RCH]{Random Channel Hopping}
\acro{rd}[RD]{Relative Direction}
\acro{relu}[ReLu]{Rectified Linear Unit}
\acro{rfc}[RFC]{Request for Comments}
\acro{rlc}[RLC]{Radio Link Control}
\acro{rl}[RL]{Reinforcement Learning}
\acro{rlink}[RL]{Reverse Link}
\acro{rnn}[RNN]{Recurrent Neural Network}
\acro{rpl}[RPL]{IPv6 Routing Protocol for Low-Power and Lossy Networks}
\acro{rr}[RR]{Round Robin} 
\acro{rrm}[RRM]{Radio Resource Management} 
\acro{rssi}[RSSI]{Received Signal Strength Indicator}
\acro{rsvp}[RSVP]{Resource Reservation Protocol}
\acro{rt}[RT]{real~time}
\acro{rtt}[RTT]{Round Trip Time}
\acro{rtx}[RTX]{Retransmit}
\acro{rx}[RX]{Receive}
\acro{sesar}[SESAR]{Single European Sky ATM Research}
\acro{sf0}[SF0]{Scheduling Function 0}
\acro{sf1}[SF1]{Scheduling Function 1}
\acro{sfg}[SFG]{Signal Flow Graph}
\acro{sfl}[SFL]{Slot Frame Length}
\acro{sfsb}[SFSB]{Scheduling Function with Soft Blacklisting}
\acro{sf}[SF]{Scheduling Function}
\acro{sfx}[SFX]{6TiSCH Experimental Scheduling Function}
\acro{sinr}[SINR]{Signal-to-Interference-plus-Noise Ratio}
\acro{snr}[SNR]{Signal-to-Noise Ratio}
\acro{sotdma}[SOTDMA]{Self-organized TDMA}
\acro{sr}[SR]{Speed Ratio}
\acro{ssr}[SSR]{Secondary Surveillance Radar}
\acro{su}[SU]{Secondary User}
\acro{svm}[SVM]{Support Vector Machine}
\acro{sw}[SW]{Stop-and-Wait}
\acro{tdm}[TDM]{Time Division Multiplexing}
\acro{tdma}[TDMA]{Time-Division Multiple Access}
\acro{tg}[TG]{Task Group}
\acro{tsch}[TSCH]{Time Slotted Channel Hopping}
\acro{tsn}[TSN]{Time-Sensitive Networking}
\acro{ts}[TS]{Time Slot}
\acro{tti}[TTI]{transmission time interval}
\acro{ttl}[TTL]{Time To Live}
\acro{tx}[TX]{Transmit}
\acro{uat}[UAT]{Universal Access Transceiver}
\acro{uav}[UAV]{unmanned aerial vehicle}
\acro{ue}[UE]{User Equipment}
\acro{ul}[UL]{uplink}
\acro{urllc}[URLLC]{Ultra-Reliable Low-Latency Communications}
\acro{v2i}[V2I]{Vehicle-to-Infrastructure}
\acro{v2v}[V2V]{Vehicle-to-Vehicle}
\acro{vae}[VAE]{Variational Auto Encoder}
\acro{vdl}[VDL]{VHF Data Link}
\acro{vhf}[VHF]{Very high frequency}
\acro{voip}[VoIP]{Voice-over-IP}
\acro{w3c}[W3C]{World Wide Web Consortium}
\acro{waic}[WAIC]{Wireless Avionics Intra-Communication}
\acro{wg}[WG]{Working Group}
\acro{wifi}[WiFi]{IEEE 802.11 family of wireless technologies}
\acro{wlan}[WLAN]{Wireless Local Area Network}
\acro{wlog}[WLOG]{without loss of generality}
\acro{wrc}[WRC]{World Radiocommunication Conference}
\acro{wsn}[WSN]{Wireless Sensor Network}

\end{acronym}

\newcommand{\SIb}[2]{\SI[number-math-rm = \mathnormal, parse-numbers = false]{#1}{\color{gray}\left[#2\right]}}
\newcommand{\omnet}{OMNeT++}

\title{A Tutorial on Trace-based Simulations of Mobile Ad-hoc Networks on the Example of Aeronautical Communications}

\author{	Musab Ahmed Eltayeb Ahmed
\and
    Konrad Fuger
\and
	Sebastian Lindner
\and
	Fatema Khan
\and
	Andreas Timm-Giel
}

\institute{
  Hamburg University of Technology,
  Hamburg, Germany\\
  \email{\{musab.ahmed, k.fuger, sebastian.lindner, fatema.khan, timm-giel\}@tuhh.de}
 }

\authorrunning{Ahmed, Fuger, Lindner, Khan and Timm-Giel}

\titlerunning{Trace-based Simulations of MANETs}

\begin{document} 

\maketitle

\begin{abstract}
  The \omnet{} simulator is well-suited for the simulation of randomized user behavior in communication networks.
  However, there are scenarios, where such a random model is unsuited to evaluate a communication system, and this paper attempts to highlight such a case.
  Using this example of ad-hoc communication between aircraft mid-flight, a tutorial-style description is attempted that shall show how the \omnet{} simulator can be used when a wealth of real-world trace data is available.
  In particular, it is described how mobility trace files can be directly used within \omnet{}, and how to link the generation of data messages to this mobility data. This is explained via an example simulation that evaluates a communication network in which an aircraft notifies the ground control when it enters or leaves a specific geographic region. Additionally, a novel trace-based application has been developed to achieve this link between mobility and message generation.
  Furthermore, a new TDMA-based medium access protocol for decentralized communication networks is presented, which is oracle-based and thus allows a TDMA-like behavior of medium access without causing any overhead; it can be useful when upper-layer protocols should be evaluated under the assumption of TDMA-like behavior, but isolated from the effects of a full-fledged TDMA protocol.
  Finally, physical layer behavior is often either overly simplistic or overly computationally expensive.
  For the latter case, when a detailed channel model is available but its evaluation requires prohibitive computational effort, then averaging its behavior into trace data can find a middle ground between efficient evaluation and realistic representation.
  Hence, a novel trace-based radio model has been developed that makes use of an SNR to PER mapping.
  In the spirit of open science, all implementations have been made available under open licenses~--~please see the conclusion.
\end{abstract}

\section{Introduction}\label{sect:introduction}
	Most of the time, when communication systems should be evaluated through simulation concerning some \ac{kpi}, randomized behavior is modeled.
	This randomization should be bounded to realistic limits so that average cases are simulated most of the time, and edge cases are investigated sometimes. When such a simulation model has been set up, it can be run many times to obtain statistically meaningful results.
	For example, when the performance of an IEEE 802.11 WiFi protocol should be studied, then it may be sufficient to model web browsing behavior through statistical means such as a negative exponentially distributed \quote{reading time} that passes before the next request is triggered by a user.
	
	Sometimes, however, specialized communication systems are tailored to specific applications and requirements, where randomized behavior does not reflect a realistic utilization of the system.
	The authors of this paper are involved in a German national research project, where the upcoming \ac{ldacs} is extended by \ac{a2a} functionality; the \ac{a2g} component of \ac{ldacs} is being standardized by \ac{sesar} and \ac{icao} at the time of writing.
	The details of \ac{ldacs} are out of the scope of this contribution, but the interested reader may refer to \cite{schnellLDACSFutureAeronautical2014}, \cite{bellido-manganellModernAirtoAirCommunications2019} for an introduction.
	In this particular case, \ac{a2a} communication involves \ac{atc} and safety-related communication.
	With this focus, the requirements concerning latency and reliability are fixed, and message generation is clearly defined.
	For example, when an aircraft enters or leaves an \ac{oca}, it may be required to communicate this event to the authorities. Other examples include requesting to change altitude, or requesting permission to land.
	These events are thus \emph{bound} to mobility, which in turn is tightly controlled. Moreover, the busiest oceanic airspace~--~i.e., the most interesting geographic area for \ac{ldacs} \ac{a2a}~--~is the \ac{nac}, as discussed in \cite{medinaFeasibilityAeronauticalMobile2008}, \cite{bellido-manganellFeasibilityFrequencyPlanning2021}.
	Here, flight routes are pre-assigned by the authorities to follow tracks, which are updated daily to accommodate up-to-date weather conditions and wind streams.
	In consequence, there exist vast amounts of mobility data of actual flight path traces, and communication data message generation can be directly linked to each aircraft and its mobility.
	
	With such a special-purpose communication system and such pre-determined mobility and data traffic in mind, randomized user behavior is not the right model to evaluate this system's \acp{kpi}.
	Instead, the need arises to use existing mobility and data traffic traces, so that real-world scenarios can be evaluated.
	In this paper, we aim to show how the \omnet{} simulator can be used to effectively make use of trace data, so that future research with these specific requirements may benefit.
	In particular, Sec.~\ref{sect:application} shows how data traffic generation can be linked to mobility, in Sec.~\ref{sect:mac} how an idealized \ac{tdma} \ac{mac} can be implemented, and how a custom radio model can reflect \ac{snr} behavior based on data that has a priori been generated using a sophisticated channel model for the respective communication system.
	Sec.~\ref{sect:results} provides the evaluation of an example scenario, where all system model components are present, as aircraft must communicate entering and leaving a particular \ac{oca} to a ground station so that messages are linked to user mobility.
	Finally, Sec.~\ref{sect:conclusion} provides a discussion and a conclusion.	
\section{System Model}\label{sect:model}
	This section is concerned with a description of the system model as well as details on the implementations that realize the respective components.
	
	\subsection{Trace-based Mobility}\label{sect:mobility}
		Trace-based mobility is, for our needs, sufficiently implemented in \omnet{} through \texttt{BonnMotionMobility}.
		Aircraft mobility traces are converted to this format, which specifies in each line the coordinates $(x,\, y,\, z)$ of a specific aircraft for each moment in time $t$.
	
	\subsection{Trace-based Data Traffic}\label{sect:application}
		As motivated earlier, data traffic generation is linked to mobility.
		For aircraft, certain conditions may trigger the generation of a message, such as entering an \ac{oca}, a different flight phase, etc.
		To achieve this, a novel \texttt{UdpTraceBasedApp} inherits from \texttt{UdpBasicApp}, which requires an additional \texttt{traceFile} configuration parameter.
		The original \texttt{UdpBasicApp}'s \texttt{startTime}, \texttt{stopTime} and \texttt{sendInterval} parameters are ignored.
		Instead, it expects a \texttt{.csv}-formatted text file that contains a simulation time stamp in each line.
		These time stamps should be monotonically increasing and specify the moment in time where a new message is generated.
		The interval inbetween messages is therefore provided through this trace file, which should be generated jointly with the mobility trace files to link mobility to message generation.
		This design allows the greatest flexibility, as different applications~--~e.g., reporting to have entered or left an \ac{oca} as opposed to requesting permission to land~--~may have different destinations or message sizes.
		Therefore, each application would be realized through their own \texttt{UdpTraceBasedApp} and corresponding trace files.
	
	\subsection{Idealized \ac{tdma} Medium Access}\label{sect:mac}
		The design of \ac{ldacs} \ac{a2a} presents several challenges.
		Firstly, the large communication ranges in the order of hundreds of kilometers prevent \ac{csmaca}-like \ac{mac} protocols, as radio wave propagation delays exceed \SI{1}{\milli\second}, so that sensing results are outdated at the time of transmission.
		The actual \ac{mac} protocol for \ac{ldacs} \ac{a2a} is therefore a \ac{tdma}-based protocol~--~just as \ac{ldacs} \ac{a2g} \cite{eppleOverviewInterferenceSituation2011}, \cite{eppleModelingDMEInterference2012}.
		Its intricacies will be published at another time, while some particular challenges have already been addressed \cite{lindnerCoexistenceSharedSpectrumRadio2020}, \cite{fisserPredictiveSchedulingOpportunistic2020}.
		
		Secondly, routing plays a major role in \ac{ldacs} \ac{a2a}, as aircraft en route over the \ac{nac} will communicate with ground stations outside of their communication ranges i.e., messages must be forwarded by intermediate aircraft until the ground network can be reached.
		To evaluate routing protocols before the proposed \ac{mac} protocol was sufficiently specified and implemented, and to focus on the effects of routing protocols in isolation~--~without superimposed effects of an underlying \ac{mac} protocol~--~an \emph{idealized} \ac{tdma} protocol was required.
		In a sense, it can be seen as a baseline \ac{mac} protocol implementation for time-slotted communication.
		\ac{ldacs} \ac{a2a} provides self-organized communication to users of a \ac{manet} with no central coordinator, which means that an allocation of time slots to users is a non-trivial task.
		The idealized \ac{tdma} implementation, on the other hand, allocates time slots to users unrealistically through an oracle: whenever a user requires a transmission slot, it is assigned an exclusive slot without any control overhead.
		When the focus of evaluation is not the \ac{mac} protocol, then this idealized \ac{tdma} \ac{mac} can be used to automatically assign time slots to users and thus achieve the characteristics of a \ac{tdma} system.
		In some \ac{manet} scenarios such as \ac{ldacs} \ac{a2a} this may be preferred over \ac{csmaca}, and it had not been implemented for \omnet{}.
		
\begin{figure}[htb]
			\begin{centering}
				\includegraphics[width=.7\textwidth]{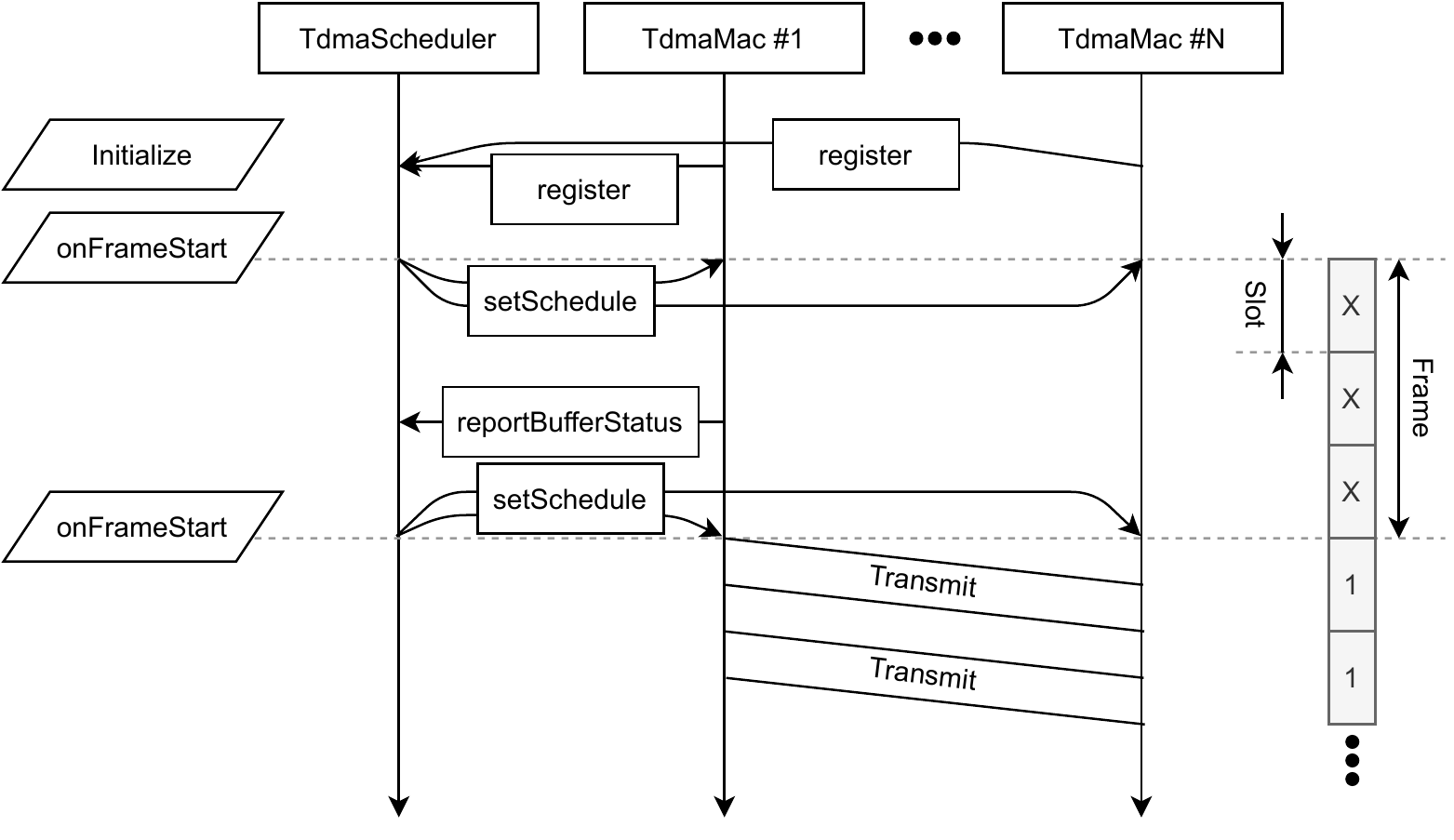}
				\caption{Flow chart of the provided idealized \ac{tdma} \ac{mac} protocol.}
				\label{fig:tdmaFlowChart}
			\end{centering}
		\end{figure}

		This has been realized through a global \texttt{TdmaScheduler} entity.
		Users are equipped with a custom \ac{nic} that provides a \texttt{TdmaMac} sublayer.
		The \texttt{TdmaScheduler} then operates as shown in Fig.~\ref{fig:tdmaFlowChart}: upon simulation start, all users with a \texttt{TdmaMac} sublayer register themselves at the \texttt{TdmaScheduler}.
		Whenever a \texttt{TdmaMac} has data to transmit, it notifies the \texttt{TdmaScheduler} about its current buffer status, which is the number of packets it is currently waiting to transmit.
		The \texttt{TdmaScheduler} is configured with a \ac{tdma} frame size; at the start of a frame, its schedule is computed.
		In the provided implementation, slots within a frame are allocated to those users that have reported a communication need in a round-robin fashion, while more sophisticated schedule computations are easily realized, which could take fairness over time or throughput into account.
		The computed schedule is then distributed to the users' \texttt{TdmaMac}s, which consequently schedule their respective transmissions.
	
	\subsection{Trace-based Radio Model}\label{sect:phy}
		The \omnet{} simulator provides a multitude of \ac{phy} layer radio models already, from simple \texttt{UnitDiskRadio} to more advanced models with modulation implementations.
		However, the realistic modeling of the radio channel can become significantly more complicated than the provided implementations.
		In fact, such channel models can require such computational effort that the evaluation of \emph{just} this model for communication between \emph{just} two users in many different constellations takes considerable time.
		When this is the case, then evaluating this model for the simulation of hundreds of users as they traverse the \ac{nac} over the course of real-world hours is out of the question due to the sheer computational complexity.
		
		In this scenario, it is preferable to evaluate the channel model for communication over water (which reflects radio waves differently than ground) once, average the results and use this simplified and averaged model to evaluate above-\ac{phy} layer protocols.
		To achieve this, a novel \texttt{TraceBasedRadio} model has been implemented, which extends the \texttt{UnitDiskRadio} model.
		Its corresponding trace file maps \ac{snr} to a priori computed \ac{per}.
		With this, arbitrarily complicated channel models can be described using a lookup table; admittedly, this decreases the model's precision, but such a channel model is usually evaluated using other means before this step.
		It should be noted that no interpolation between the given $\text{\ac{snr}} \rightarrow \text{\ac{per}}$ values is done, as the model has been found to behave non-linearly and no fitting interpolation could be identified.
		The trace file should therefore contain an adequate resolution of these mappings, as the implementation looks for the closest match.
		
		This model is obtaining the \ac{snr} similarly to what is done in \cite{bellido-manganellFeasibilityFrequencyPlanning2021}. The radio horizon $r_\text{h}(h_\text{tx},h_\text{rx})$ defines the farthest possible point of propagation and it is given as in Eq.~\ref{eq:radio-horizon}, where $h_\text{tx},\, h_\text{rx}$ are the heights of transmitter and receiver respectively.
		\begin{equation}\label{eq:radio-horizon}
			\SIb{r_\text{h}(h_\text{tx},\, h_\text{rx})}{\kilo\meter} = 130.4(\sqrt{\SIb{h_\text{tx}}{\kilo\meter}} + \sqrt{\SIb{h_\text{rx}}{\kilo\meter}}).
		\end{equation}
		The \ac{fspl} denoted by $L_p(d, f, h_\text{tx}, h_\text{rx})$ is then found as in Eq.~\ref{eq:fspl}, where $d$ is the distance between the transmitter and the receiver and $f$ is the transmission frequency; we assume that $L_\text{p} \rightarrow \infty$ beyond the radio horizon.
		\begin{equation}\label{eq:fspl}\small
			\SIb{L_p(d, f, h_\text{tx}, h_\text{rx})}{\decibel} = \begin{cases} 20\log_{10} (\SIb{d}{\kilo\meter}) + 20\log_{10} (\SIb{f}{\mega\hertz}) + \SI{32.4478}{\color{gray}[\dB]}, \quad \text{if } d < r_\text{h}. \\ \infty, \quad \text{else}. \end{cases}
		\end{equation}	
		The received power $P_\text{rx}$ is then found from Eq.~\ref{eq:rcvd-power}, where $P_\text{tx}$ is the transmission power, $G_\text{tx}$ and $G_\text{rx}$ are the transmitter and receiver gains respectively and $L_\text{tx}$ and $L_\text{rx}$  represent the transmitter and receiver losses respectively.
		\begin{equation}\label{eq:rcvd-power}
			\SIb{P_\text{rx}}{\deci\belmilliwatt} = \SIb{P_\text{tx}}{\deci\belmilliwatt} + \SIb{G_\text{tx}}{\dBi} - \SIb{L_\text{tx}}{\dB} + \SIb{G_\text{rx}}{\dBi} - \SIb{L_\text{rx}}{\dB} - \SIb{L_p(d, f, h_\text{tx}, h_\text{rx})}{\decibel}
		\end{equation}
		
		Finally, the \ac{snr} of a communication link is found using Eq.~\ref{eq:snr}, where $F_N$ is the noise figure, $N_0$ is the thermal noise density and $B$ is the receiver bandwidth.
		\begin{equation}\label{eq:snr}
			\SIb{\text{\ac{snr}}}{\dB} = \SIb{P_\text{rx}}{\dBm} - (\SIb{F_N}{\dB} + \SIb{N_0}{{\frac{\dBm}{\Hz}}} + 10\log_{10} \SIb{B}{\Hz})								
		\end{equation}
		
\begin{figure}[htb]
		\begin{centering}
			\includegraphics[width=0.6\textwidth]{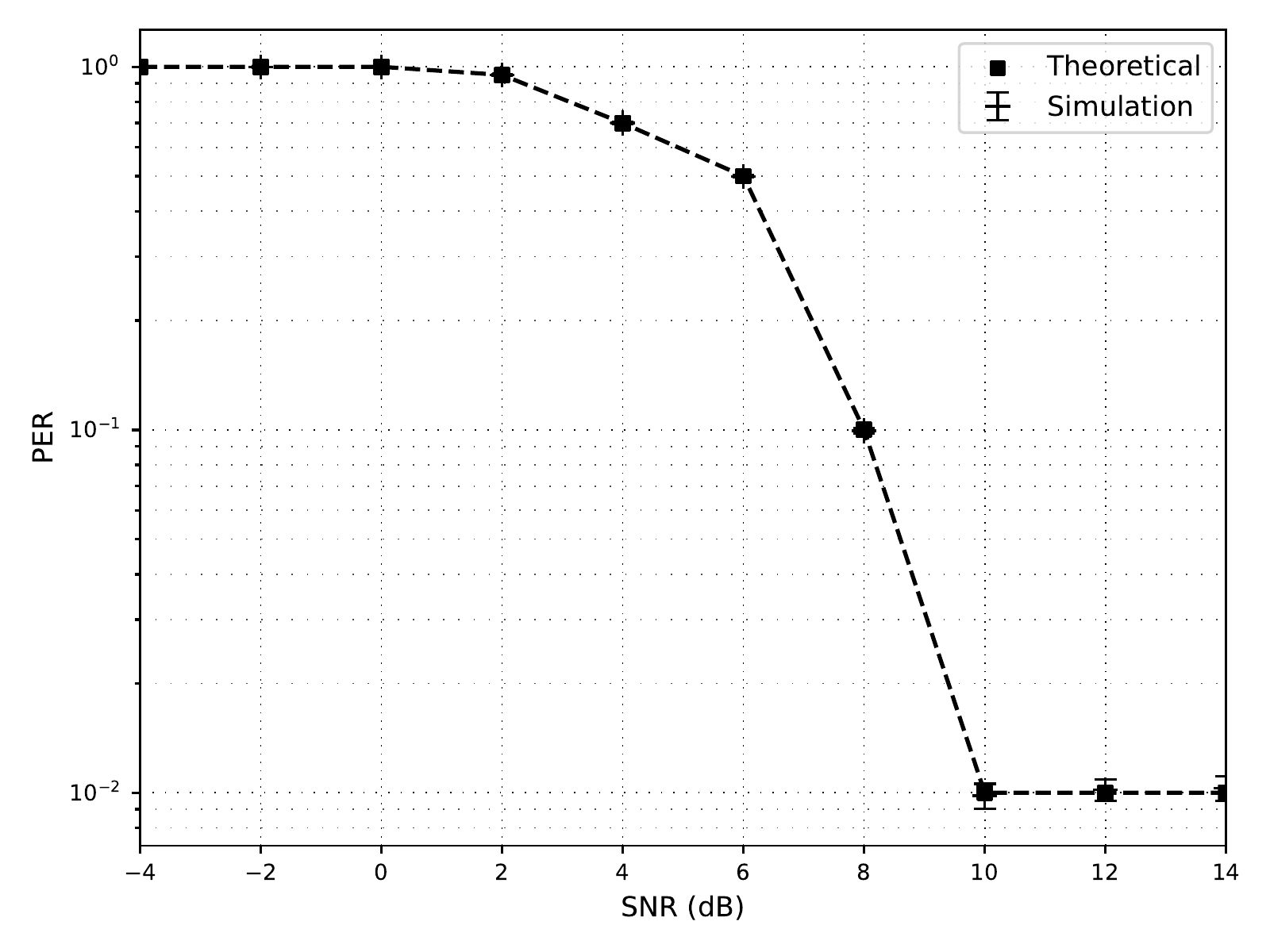}
			\caption{A validation scenario compares expected \ac{per} to the one observed through simulation using the provided trace-based radio model.}
			\label{fig:snr-vs-per}
		\end{centering}
	\end{figure}

		A provided \texttt{TraceBasedReceiver} extends the \texttt{UnitDiskReceiver} model by performing this computation, and passing the \ac{snr} to a novel \texttt{TraceBasedErrorModel}, which queries the trace file-provided lookup table to obtain the closest-matching \ac{snr}'s mapping to a \ac{per}.
		This trace file is expected to contain a triple of (\ac{snr}, \ac{per}, \ac{ber}) per line.	
		A validation scenario positioned the receiver at $(x=\SI{0}{\kilo\meter}, y=\SI{0}{\kilo\meter}, z=\SI{30}{\kilo\meter})$, and the transmitter at $(x \in \{180, 220, 275, \ldots{}, 1400\} \, \text{km}, y=\SI{0}{\kilo\meter}, z=\SI{30}{\kilo\meter})$.
		The distances are chosen to reflect the entire range from certain success to certain failure, and Fig.~\ref{fig:snr-vs-per} shows that simulated results are as expected.
		The variance in simulation results stems from the fact that the lookup table returns the probability of successfully receiving a packet; a random number is then drawn to determine the actual outcome of this transmission attempt, and so the observed \ac{per} meets the trace file-provided one on average with some small variance.
		
\section{Results}\label{sect:results}
\begin{figure}[htb]
			\begin{centering}
				\includegraphics[width=0.6\textwidth]{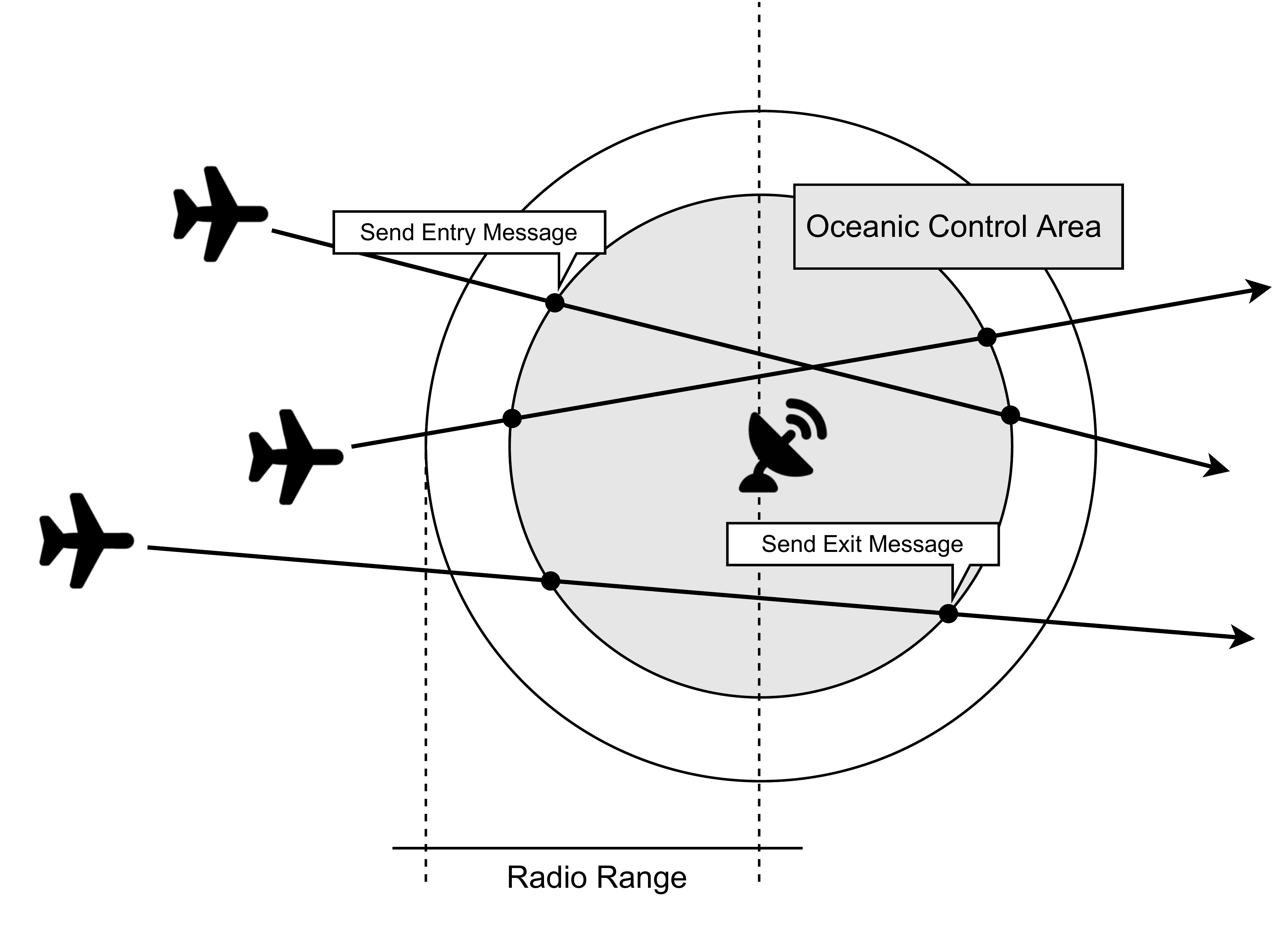}
				\caption{Evaluation Scenario}
				\label{fig:scenario_schema}
			\end{centering}
		\end{figure}

		An exemplary evaluation scenario is visualized in Fig.~\ref{fig:scenario_schema}.
		It aims to incorporate all previously described aspects into one simulation model, so that an adaption to other scenarios is easily achieved.
		In it, a particular \ac{oca} is controlled by a ground station positioned in its center.
		Upon entering or leaving the \ac{oca}'s range, all aircraft are required to communicate the respective event to the ground station with some pre-defined control message.		
		Simulations have been performed using OMNeT++ v5.6.2 and INET v4.2.5.	
		The simulation parameters are given in Table~\ref{tab:ltbexample}.
	
		\begin{table}[htb]
			\begin{centering}
				\begin{tabular}{|l|l|}
					\hline
					Number of users             & $n \in \{100, 200, \ldots, 500\}$ \\ \hline
					\ac{mac}             & Idealized \ac{tdma}         \\ \hline
					TDMA slot duration   & $\SI{10}{\milli\second}$                \\ \hline
					TDMA number of slots & $\SI{10}{\text{slots}}$                   \\ \hline
					TDMA retransmission attempts & $0$                   \\ \hline
					Radio model          & \texttt{traceBasedRadioModel} \\ \hline
					Radio range        & $\SI{400}{\kilo\meter}$           \\ \hline
					\ac{oca} range        & $\SI{370.4}{\kilo\meter}$           \\ \hline					
					Number of runs       & $10$                   \\ \hline
					Simulation time      & $\SI{10000}{\second}$             \\ \hline
				\end{tabular}
				\caption{Simulation parameters of the exemplary aeronautical \ac{manet} scenario.}
				\label{tab:ltbexample}
			\end{centering}
		\end{table}
		
\begin{figure}[htb]
			\begin{centering}
				\includegraphics[width=.6\textwidth]{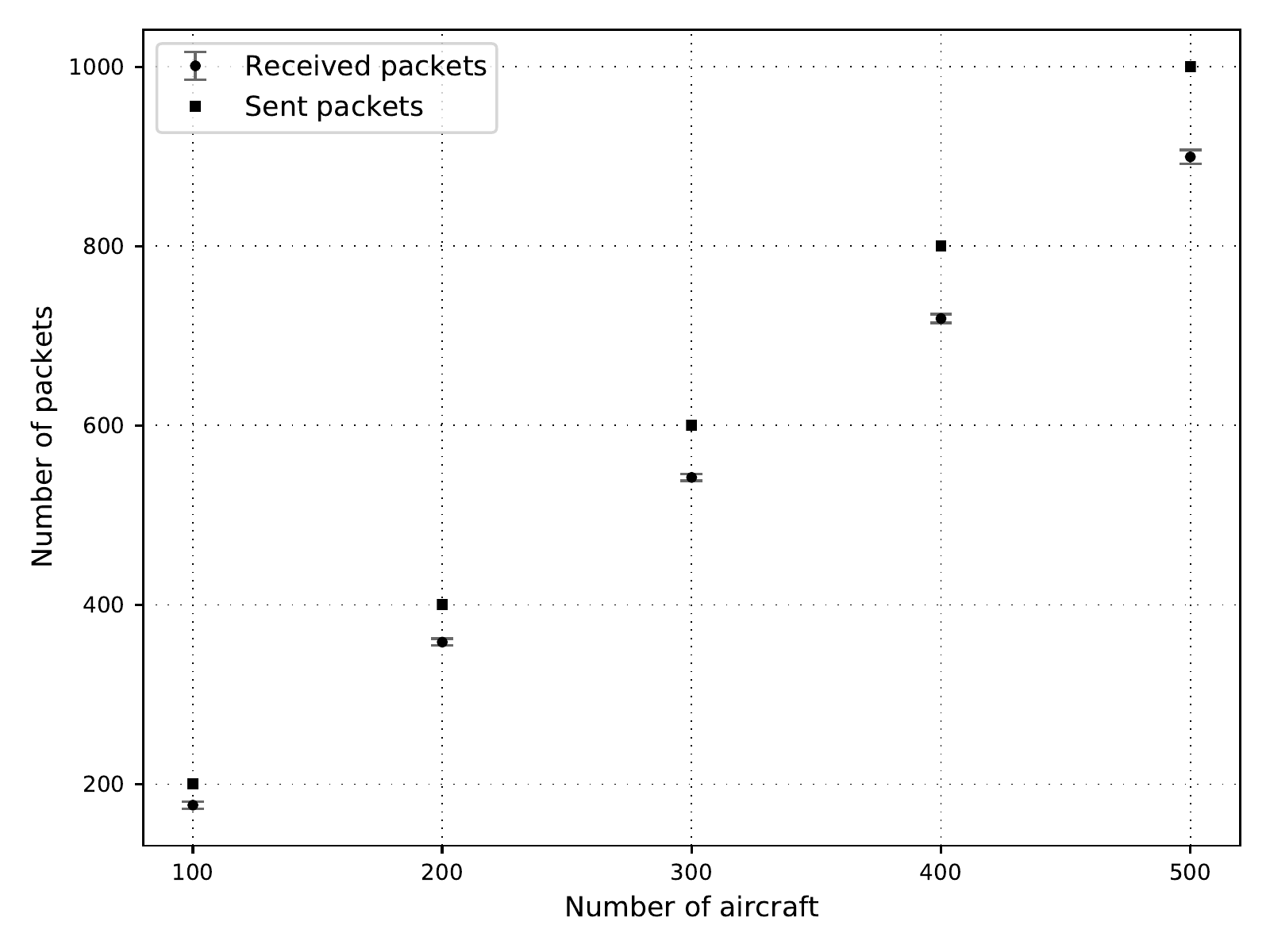}
				\caption{The number of received packets with different number of aircraft considered in the scenario.}
				\label{fig:per}
			\end{centering}
		\end{figure}

		The number of received packets at the ground station is evaluated in Fig.~\ref{fig:per}.
		As would be expected, the number of \emph{sent} packets is exactly twice the number of users: one packet for \ac{oca} entry and one for \ac{oca} exit.
		The number of \emph{received} packets is $\SI{10}{\percent}$ smaller, which is due to the chosen \ac{oca} range at $\SI{370.4}{\kilo\meter}$, which translated to a \ac{snr} of \SI{8}{\deci\bel} and thus a \ac{per} of \SI{10}{\percent} as shown in Fig.~\ref{fig:snr-vs-per}, and which explains the gap between sent and received packets of $\SI{10}{\percent}$ in Fig.~\ref{fig:per}.
		
\section{Conclusion}\label{sect:conclusion}
	Trace-based simulation has, in this paper, been motivated by the evaluation of \ac{ldacs} \ac{a2a}.
	In this case, abundant real-world data is available, which presents the opportunity of simulating a communication system under realistic circumstances.
	In particular, \ac{ldacs} \ac{a2a} is going to be utilized for \ac{atc} and safety-related applications, where message generation is clearly defined through trigger events.
	We have provided a tutorial-style summary of how such a scenario can be simulated.
	Real-world mobility data presents the basis, to which a novel data traffic application can bind the generation of messages.
	An idealized \ac{tdma} \ac{mac} protocol fills a gap in the \omnet{} toolbox, as it realizes a \ac{tdma} communication system suitable for simulations where effects of a full-fledged \ac{mac} should be suppressed.
	With its oracle-based design, it can be used even for \acp{manet}, where no central coordinator is present, and for aeronautical networks, where propagation delays prevent \ac{csmaca}-like approaches.
	The necessity of including a complicated radio channel model required the implementation of a trace-based radio model for the \omnet{} simulator.
	It averages channel behavior by mapping \ac{snr} to \ac{per}, and so provides an efficient integration of behaviors that usually take considerable computational effort to model.\\	
	The presented evaluation scenario binds all of these aspects together.
	First, the mobility-determined event of entering or leaving an \ac{oca} triggers the generation of corresponding control messages.
	Moreover, the idealized \ac{tdma} allocates transmission slots to users and so prevents interference.
	Additionally, the trace-based radio model has been shown to model a priori defined \acp{per} accurately.\\
	All implemented modules are available under open licenses.
	They are released in such a way that future research into trace-based scenarios benefits, as the simulation model and components are easily adaptable.
	The trace-based data traffic application is implemented in \cite{eltayebahmedCodeReleaseUdpTraceBasedApp2021}, the \ac{tdma} module is implemented in \cite{fugerCodeReleaseTDMA2021}, the radio model in \cite{khanCodeReleaseTrace2021} and the example simulation model in \cite{eltayebahmedCodeReleaseTracebased2021}.
	The exemplary scenario was chosen in such a way that it is easy to understand as well as verify.
	In future work, the trace-based radio model should support \ac{ber} instead of just the \ac{per}, and it should work with \ac{sinr} instead of just \ac{snr}.
	In particular, an extension to a radio model that understands multiple, orthogonal frequency channels is foreseen, where simultaneous transmissions within \emph{one} channel should cause interference upon each other, but transmissions in \emph{different} channels should not.

\section{Acknowledgments}\label{sect:acks}
	This work was partially funded by the German Federal Ministry for Economic Affairs and Energy (BMWi) as part of the \textit{IntAirNet} project with reference number 20V1708F.

\appendix
\renewcommand*{\bibfont}{\small}
\printbibliography

\end{document}